\documentclass[%
 reprint,
 amsmath,amssymb,
 aps,
]{revtex4-2}

\usepackage{graphicx}
\usepackage{dcolumn}
\usepackage{bm}

\usepackage{graphicx}	
\usepackage{amsmath}	
\usepackage{amssymb}	
\usepackage{xcolor}
\usepackage{soul}

\def\bu{\boldsymbol{u}}
\def\bB{\boldsymbol{B}}

\def\bk{\boldsymbol{k}}

\def\muQ{\mu_{\mathcal{Q}}}
\def\tQ{t_{\mathcal{Q}}}
\def\BQ{B_{\mathcal{Q}}}
\def\Q{\mathcal{Q}}

\def\Gf{\Gamma_{\mathrm{f}}}

\def\Eq{Eq.}
\def\Eqs{Eqs.}
\def\beq{\begin{equation}}
\def\eeq{\end{equation}}
\def\tt{\widetilde{t}}
\def\ttsat{\widetilde t_{\mathrm{sat}} }
\def\tp{\widetilde{p}}
\def\tL{\widetilde{L}}
\def\tB{\widetilde{B}}
\def\tbB{\widetilde{\bB}}
\def\tbu{\widetilde{\bu}}
\def\tmu{\widetilde{\mu}}
\def\tk{\widetilde{k}}
\def\tbk{\widetilde{\bk}}
\def\tnabla{\widetilde{\nabla}}
\def\tS{\widetilde{S}}
\def\tgamma{\widetilde{\gamma}}
\def\tHtot{\widetilde{{\cal H}}}
\def\tH{\widetilde{H}}
\def\Pm{{\rm Pm}}
\def\Rm{{\rm Rm}}
\def\Htot{{\cal H}}

\begin{document}

\preprint{APS/123-QED}

\title{Inefficiency of chiral dynamos in protoneutron stars and the early Universe}
\author{Valentin A. Skoutnev}
\affiliation{Physics Department and Columbia Astrophysics Laboratory, Columbia University, 538 West 120th Street New York, New York 10027,USA}
\email{valentinskoutnev@gmail.com}
 
\author{Andrei M. Beloborodov}
\affiliation{Physics Department and Columbia Astrophysics Laboratory, Columbia University, 538 West 120th Street New York, New York 10027,USA}
\affiliation{Max Planck Institute for Astrophysics, Karl-Schwarzschild-Str. 1, D-85741, Garching, Germany}
\date{\today}

\begin{abstract}
The chiral plasma instability (CPI) has been invoked as a possible mechanism for generating primordial magnetic fields in the Universe and ultrastrong fields in neutron stars. We investigate chiral dynamos where the chirality imbalance is pumped by a source on a timescale $t_0$ and show that the CPI rate $\gamma$ is limited to $\gamma_0/(1+\Q^2)$, where $\Q= (\gamma_0 t_0)^{1/3}$ and $\gamma_0$ corresponds to models with instantaneously created chirality imbalance $(t_0=0)$. We then find that chiral flipping with rate $\Gf$ hinders the chiral dynamo if $\Gf >\gamma_0/(1+\Q^2)$ and completely suppresses it if $\Gf >\gamma_0/(1+\Q^{3/2})$. Realistic $t_0$ typically give $\Q\gg 1$, which makes  the dynamo greatly vulnerable to suppression by chiral flipping. The suppression is strong in protoneutron stars and may be (barely) avoided near the electroweak transition in the early Universe.
\end{abstract}

\maketitle

\section{Introduction}

The amplification of helical magnetic fields in relativistic astrophysical plasmas may be linked with the chiral anomaly of weak interactions (see \cite{kamada2023chiral} for a review). A plasma with a chirality imbalance has unequal densities $n_R\neq n_L$ of right and left handed electrons (i.e. electrons with spins aligned and anti-aligned with their momentum, respectively). The density imbalance $n_5=n_R-n_L$ implies imbalance in chemical potentials $\mu_5=\mu_R-\mu_L$. This provides an energy reservoir (with density $\sim n_5\mu_5$) that can be tapped by the chiral plasma instability (CPI) to amplify helical magnetic fields. The CPI respects a topological invariant and converts particle chirality into magnetic helicity without changing their sum. However, $\mu_5$ is also irreversibly reduced by ``chiral flipping" due to particle collisions. Magnetic field generation by the CPI may fail if chirality imbalance is converted into magnetic helicity slower than it is destroyed by chiral flipping.

The CPI has been invoked as a potential origin of primordial magnetic fields, relying on the chirality imbalance that may be present in the early Universe \citep{joyce1997primordial,frohlich2000new,boyarsky2012self,tashiro2012chiral,pavlovic2016modified,brandenburg2017turbulent,schober2018laminar,schober2022production,gurgenidze2025primordial}. It was considered for the early expansion phase with temperature $k_{\mathrm{B}}T\gtrsim80$\,TeV (when chiral flipping rates were longer than the Hubble time) and near the electroweak phase transition at $k_{\rm B}T\sim 100$\,GeV, assuming chirality imbalance is generated near the transition. The CPI has also been invoked to explain strong magnetic fields in protoneutron stars (PNS). In this case, the chirality imbalance develops in the process of neutron star formation: compression of the collapsing progenitor core drives capture reactions of left-handed electron onto protons ($e^-_L+p\rightarrow n+\nu^e_L$), as only left-handed electrons participate in the weak interactions. Most optimistic estimates suggested that CPI accessing the energy reservoir $n_5\mu_5\lesssim n_e\mu_e$ could generate fields with strengths ranging from $10^{14}\,$G to $10^{18}\,$G \citep{akamatsu2013chiral,ohnishi2014magnetars,dvornikov2015magnetic,sigl2016chiral,masada2018chiral,schober2018laminar,matsumoto2022chiral}, where $n_e=n_R+n_L$ and $\mu_e=(\mu_R+\mu_L)/2$ refer to the total electron population. However, chiral flipping is a crucial limiting factor for the CPI in PNS \citep{grabowska2015role,sen2026chiral}. 
 
Many previous works have investigated the CPI assuming an initially large chirality imbalance, which leads to CPI development at the maximal growth rate $\gamma_0$. However, this assumption is typically invalid in astrophysical systems where the evolutionary timescale to build up imbalance, $t_0$, is much longer than the shortest growth timescale of the CPI, $t_0\gg\gamma_0^{-1}$. Feedback by the CPI should thus occur before the imbalance becomes large. In this paper, we examine the evolution of a chiral plasma whose imbalance is driven on a finite timescale, similar to studies invoking an explicit source term to pump $\mu_5$ \citep{grabowska2015role,sigl2016chiral,pavlovic2017chiral,gurgenidze2025primordial}. We provide analytical estimates for the resulting magnetic fields (Section~\ref{sec:Theory}) and confirm them with direct numerical simulations (Section~\ref{sec:Sim}). The estimates lead to a new criterion for chiral flipping to suppress the CPI, which we evaluate for the early Universe and PNS (Section~\ref{sec:Apps}).

\section{Chiral dynamo}
\label{sec:Theory}

\subsection{Equations of Chiral Magnetohydrodynamics}
\label{sec:TheoryCMHD}

The evolution of the chiral density $n_5$ in a fluid moving with velocity $\bu$ is described by \footnote{Hereafter, we use CGS units in contrast to most papers on CPI.}
\beq
\label{eq:MHD_n5Evol}
 (\partial_t+\bu\cdot\nabla)n_5 = \frac{\alpha}{2\pi^2\hbar} \boldsymbol{E}\cdot\bB+S_5-\Gf n_5,
\eeq
where $S_5$ is a source term, $\Gf$ is the chiral flipping rate, $\boldsymbol{E}$ is the electric field, and $\boldsymbol{B}$ is the magnetic field. The term $\propto\boldsymbol{E}\cdot\bB$ is the effect of the chiral anomaly (e.g. \cite{peskin2018introduction}); $\alpha=e^2/\hbar c$ is the fine structure constant.

The density $n_5$ couples to the evolution of the magnetic field through the chiral magnetic effect \citep{vilenkin1980equilibrium},
formulated as a field-aligned electric current:
\begin{equation}
\label{eq:CME}
\boldsymbol{J}_5\equiv\frac{\alpha}{\pi\hbar} \mu_5\bB. 
\end{equation}
The current induction may be pictured as the tendency for electron spin to anti-align with $\bB$ (while its magnetic moment tends to align); a preferred handedness of the electron population ($n_5\neq 0$) then implies a preferred direction for electron motion, creating current $\boldsymbol{J}_5$.

Note that the applicability of this model to massive electrons ($m_e\neq0$) may be questioned since \Eq~(\ref{eq:MHD_n5Evol}) omits a term proportional to the electron mass, and \Eq~(\ref{eq:CME}) was also derived for massless charged fermions \citep{vilenkin1980equilibrium}. Following previous works, we proceed under the assumption that this model is valid for ultra-relativistic electrons. We wish to explore if this model can generate strong magnetic fields.

The coupled evolution of $\bB$ and $\bu$ obeys magnetohydrodynamics (MHD) with the additional current 
$\boldsymbol{J}_5$ \citep{kamada2023chiral}:
\begin{align}
    \label{eq:MHD_uEvol}
    (\partial_t+\bu\cdot\nabla)\bu
    &= -\frac{\nabla p}{\rho}+\frac{(\nabla\times\bB)\times\bB}{4\pi\rho}+\nu\, \nabla^2\bu,\\
    \label{eq:MHD_BEvol}
    \partial_t\bB&=-c\,\nabla\times\boldsymbol{E},    \\
  \frac{c\boldsymbol{E}  +\bu\times\bB}{\eta} &= \nabla\times\bB-\frac{4\pi\boldsymbol{J}_5}{c},
\label{eq:ohm}
\end{align}
where $p$ is the pressure, $\rho$ is the mass density, $\nu$ is the kinematic viscosity, and
\beq
\eta=\frac{c^2}{4\pi\sigma}
\eeq
is the magnetic diffusivity ($\sigma$ is the fluid conductivity). We will treat the fluid as incompressible with $\rho=const$, which is a good approximation when motions are subsonic. The pressure $p$ then acts as a constraint variable that enforces the incompressibility of the fluid:
\begin{align}
    \label{eq:incom}
    \nabla\cdot  \bu=0.
\end{align}

\Eqs~(\ref{eq:MHD_n5Evol})-(\ref{eq:incom}) constitute a closed set of ``chiral MHD'' equations when the relation between $n_5$ and $\mu_5$ is taken into account:
\beq
\label{eq:n5}
  n_5=\frac{\mu_5}{24\pi^2\hbar^3 c^3}\times \left\{\begin{array}{cr}
    \mu_5^2+12\mu_e^2 & \quad k_{\mathrm{B}}T\ll \mu_e 
    \\
    (2\pi k_{\mathrm{B}}T)^2  & \quad k_{\mathrm{B}}T\gg \mu_e
             \end{array}\right.
\eeq
The degenerate limit is relevant to PNS, and the non-degenerate limit is relevant to the early Universe. It will be shown below that for realistic conditions $\mu_5\ll\mu_e$, so both cases in \Eq~(\ref{eq:n5}) reduce to the form $n_5\propto\mu_5$. \Eqs~(\ref{eq:MHD_n5Evol}) and (\ref{eq:CME}) may then be rewritten as
\beq
\label{eq:MHD_muEvol}
(\partial_t+\bu\cdot\nabla)\mu=2\lambda c \boldsymbol{E}\cdot\bB+S-\Gf\mu, 
\eeq
\beq
  \boldsymbol{J}_5\equiv\frac{c}{4\pi} \mu \bB.
\eeq
where
\beq
   \mu\equiv \frac{4\alpha}{\hbar c}\mu_5, \qquad   S\equiv\frac{4\pi^2\lambda\hbar c}{\alpha}S_5,
\eeq
\beq
\label{eq:lambda}
  \lambda=\alpha^2 \hbar c \times \left\{\begin{array}{cr}
    2\mu_e^{-2} & \quad k_{\mathrm{B}}T\ll \mu_e 
    \\
    6(\pi k_{\mathrm{B}}T)^{-2}  & \quad k_{\mathrm{B}}T\gg \mu_e
             \end{array}\right.
\eeq
The new variable $\mu$ has the dimension of a wavenumber (inverse length). Hereafter $\mu$ is called ``chiral potential'' or ``chirality imbalance''.

A key conservation law can be stated for the quantity
\begin{align}
\label{eq:helicity}
{\cal H}\equiv\langle\mu\rangle+\lambda H,
\qquad H=\langle\boldsymbol{A}\cdot\bB\rangle,
\end{align}
where the brackets indicate a volume average, $\langle...\rangle=V^{-1}\int ... dV$, and $\boldsymbol{A}$ is the vector potential ($\nabla\times\boldsymbol{A}=\boldsymbol{B}$). $\cal H$ is loosely called ``total chirality'' or ``total helicity.'' The time derivative of $\langle\mu\rangle$ is governed by \Eq~(\ref{eq:MHD_muEvol}), and the time derivative of $\lambda H$ can be expressed using $\boldsymbol{E}=-\nabla \phi-c^{-1}\partial_t\boldsymbol{A}$:
\begin{align}
\label{eq:H_Evol}
\dot{H}=-2c\langle\boldsymbol{E}\cdot \bB\rangle+\langle\nabla\cdot \boldsymbol{F}_h\rangle,
\end{align}
where $\boldsymbol{F}_h=c\boldsymbol{A}\times\boldsymbol{E}-c\phi \bB$.
Then, one finds
\begin{align}
\label{eq:Hdot}
\dot{\cal H}= \langle S\rangle-\left\langle \Gf \mu\right\rangle+  V^{-1}\int_{\partial V}(\boldsymbol{F}_{\mu}+\lambda\boldsymbol{F}_h)\cdot d\boldsymbol{S}, 
\end{align}
where $ \boldsymbol{F}_\mu=-\mu\bu$, and $\partial V$ is the boundary of volume $V$. The total helicity ${\cal H}$ can only change due to a source, chiral flipping, or fluxes across boundaries. We will consider volumes for which the boundary term vanishes.

\subsection{Chiral plasma instability}
\label{sec:TheoryCPI}

The canonical CPI setup assumes an initial chirality imbalance $\mu=\mu_0$ and negligible initial magnetic helicity $\lambda H_0\ll\mu_0$ \citep{masada2018chiral,schober2020chiral,matsumoto2022chiral}. The CPI amplifies helical modes of a seed magnetic field $\boldsymbol{B}(\boldsymbol{x},t)=\Re[\sum_{\bk}\bB_{0}(\bk) \exp(i\boldsymbol{k}\cdot \boldsymbol{x}+\gamma(k) t)]$ that satisfy $i\boldsymbol{k}\times  \bB_{0}(\bk)= k  \bB_{0}(\bk)$, where $k=|\boldsymbol{k}|$. Its growth rate, $\gamma(k)=\eta k(\mu-k)$, peaks at $k=\mu/2$:
\beq
\label{eq:gamma}
  \gamma=\frac{\eta \mu^2}{4}.
\eeq
The instability converts chirality imbalance $\mu$ into magnetic helicity $\lambda H$ until they become comparable, sharing the initial chirality reservoir $\mu_0$: $\lambda H\approx\mu\approx\mu_0/2$, which corresponds to field strength $B\approx \mu_0/\sqrt{\lambda}$ since $H\approx B^2/k$. The exponential amplification of the seed field at $\mu\approx\mu_0$ lasts for a time $t\approx 0.5\gamma_0^{-1} \ln[\mu_0/\lambda E_B(\mu_0/2)]$, where $\gamma_0=\eta\mu_0^2/4$ and $E_B(k)$ is the initial magnetic energy spectrum. At times $t\gg \gamma_0^{-1}$, the magnetic fields undergo an inverse cascade, increasing their length scale $L$ and decreasing their strength $B$ while conserving magnetic helicity $H\sim L B^2$.

\subsection{Evolution with a source}
\label{sec:TheorySource}

\begin{figure}
    \centering
    \includegraphics[width=\linewidth]{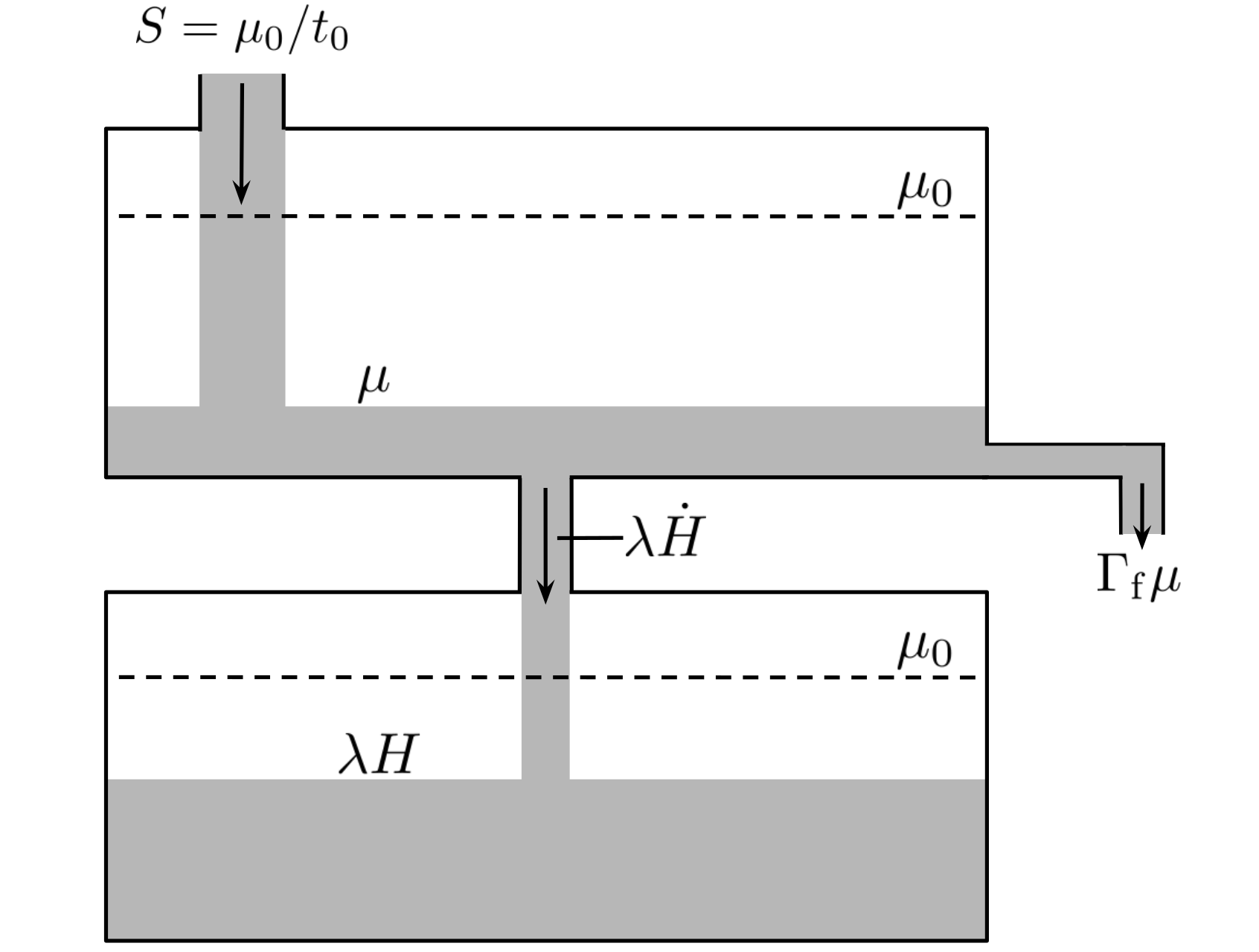}
    \caption{
    Chirality/helicity flow in a system where a source $S$ pumps the chirality imbalance $\mu$ by an amount $\mu_0$ over a timescale $t_0$. 
    The accumulated $\mu$ has two sinks: conversion into magnetic helicity by the CPI at rate $\lambda \dot H$ and destruction by chiral flipping at rate $\Gf\mu$. The amount of generated magnetic helicity $H$ is determined by the competition between the CPI and chiral flipping.
    }
    \label{fig:Schematic}
\end{figure}

A more realistic CPI setup assumes that the chirality imbalance is pumped by some source $S$ on a finite  timescale $t_0$ rather than given as an initial condition. Let us consider first the evolution without flipping, $\Gamma_{\rm f}=0$. Using \Eqs~(\ref{eq:helicity}) and (\ref{eq:Hdot}) and dropping the bracket notation for volume averages, the evolution of helicity may be stated as
\begin{align}
\label{eq:H_Evol_source_only}
\dot{\cal H}=\dot{\mu}+\lambda \dot{H}=S.
\end{align}
The source will deposit a total  helicity $ {\cal H}=\int_0^{t}S(t')dt'$, and its partition ${\cal H}=\mu+\lambda H$ depends on the strength and duration of the source, as explained below and illustrated in Fig.~\ref{fig:Schematic}.

Consider a source with the simple form,
\beq
  S(t)=\left\{\begin{array}{cr}
    \mu_0/t_0 & \;\; 0<t\leq t_0 \\
    0 & t>t_0
           \end{array}\right.
\eeq
The evolution starts with a linear growth of $\mu=St$ while the CPI growth rate (\Eq~\ref{eq:gamma}) evolves as $\gamma=\eta\mu^2/4\propto t^2$. The maximum $\mu=\mu_0$ and the maximum CPI growth rate $\gamma_0=\eta\mu_0^2/4$ are reached if $\gamma_0 t_0< 1$. In this case, the source essentially generates an initial condition with $\mu=\mu_0\gg\lambda H$ before any significant CPI occurs. The generated $\mu_0$ later relaxes through the CPI on timescale $t\sim\gamma_0^{-1}\gg t_0$ as described in Section~\ref{sec:TheoryCPI}. 

In the opposite (and realistic) regime of $\gamma_0 t_0\gg 1$, the condition $\gamma t\sim 1$ is reached before $t_0$. Then, the CPI has sufficient time to saturate and deplete a significant fraction of the pumped $\mu$, thereby quenching its growth and preventing $\mu$ from rising to $\mu_0$. The evolution toward saturation may be estimated as follows. Substituting the magnetic helicity growth rate $\dot H\sim  2\gamma H\sim (\eta \mu^2/2) H$ into \Eq~(\ref{eq:H_Evol_source_only}), one obtains
\begin{align}
    \dot\mu+\frac{1}{2}\eta \mu^2\lambda H\sim S.
\end{align}
The initial linear growth of $\mu\approx St$ is quenched when the CPI converts a significant fraction of $\mu$ into magnetic helicity: $\lambda H\sim \mu/2$. At the saturation point, $\lambda \dot H\sim\eta\mu^3/4$ is comparable to $S$, which gives $\mu\sim (4S/\eta)^{1/3}$. This estimated value of the saturated $\mu$ will be denoted as $\muQ$. Note that $\mu_0\gg\muQ$ when $\gamma_0 t_0\gg 1$. Indeed, using $S=\mu_0/t_0$, the ratio $\Q=\mu_0/\muQ$ may be rewritten as 
\beq
\label{eq:Q}
 \Q =(\gamma_0t_0)^{1/3}=(\eta\mu_0^2t_0/4)^{1/3}.
\eeq

The initial growth of $\mu\approx St$ toward $\muQ$ takes time $\tQ=t_0\muQ/\mu_0$. The exact saturation time $t_{\rm sat}$ differs from $\tQ$ by a logarithmic numerical factor that depends on the number of $e$-foldings needed to amplify the seed magnetic field to saturation. The magnetic field $B\sim (k H)^{1/2}$ at saturation is determined by $H\sim\muQ/\lambda$ and the CPI wavenumber $k\sim \muQ$. These estimates may summarized as follows
\beq
\muQ=\frac{\mu_0}{\Q}, \qquad\tQ=\frac{t_0}{\Q}, \qquad
\BQ=\frac{\muQ}{\sqrt{\lambda}}.
    \label{eq:Q_def}
\eeq

At times $t\gg\tQ$ and before $t_0$, we expect that the system finds a steady state where $\mu$ stays approximately constant as its pumping by the source is balanced by its conversion into magnetic helicity by the CPI. The steady state $\dot\mu\approx0$ and $\lambda\dot H\approx S$ gives
\beq
\label{eq:linear_growth}
  \mu\approx \muQ, \qquad \lambda H\approx S t-\muQ, \qquad
\tQ\ll t<t_0.
\eeq
The accurate saturated value of the chiral imbalance $\mu$ differs from $\muQ$ by a numerical factor of order unity. Exact calculations will take into account that the balance $\lambda\dot H\approx S$ is affected by the redistribution of magnetic helicity $H$ toward larger scales through a nonlinear inverse cascade. The accurate saturated value of $\mu$ will be found with 
numerical simulations (see Section~\ref{sec:SimSource}).

\Eq~(\ref{eq:linear_growth}) demonstrates that the chiral dynamo with $\Q\gg 1$ gives practically full conversion of $\mu_0$ into $\lambda H$ over the pumping timescale $t_0\gg\gamma_0^{-1}$: the final $\lambda H\approx \mu_0-\muQ\approx \mu_0$ \footnote{Generation of $\lambda H$ occurs linearly with time for a constant source $S$, and for a more general source $S(t)$ one would find $\lambda H(t)\approx \int_0^t S(t')\,dt'$.}. 
These dynamics significantly differ from those of a chiral dynamo with $\Q\ll 1$, where $H$ grows exponentially fast on the timescale $\gamma_0^{-1}$.

\subsection{Evolution with chiral flipping}
\label{sec:TheoryFlipping}
The helicity evolution equation with chiral flipping is 
\begin{align}
\label{eq:flip}
    \dot{\cal H} =\dot\mu+\lambda \dot H=S-\Gf\mu.
\end{align}
In the regime of $\Q\ll 1$ (which was mainly studied before), the source immediately generates $\mu=\mu_0$, and then the CPI converts it into magnetic helicity with rate $\gamma_0$. This process is not hindered by flipping if $\Gf\ll \gamma_0$ \citep{schober2018laminar}. 

Let us now consider the regime of $\Q\gg1$. The production of magnetic helicity via CPI occurs with $\mu\approx\muQ$ if chiral flipping has a negligible effect, $\Gf\muQ\ll S$. This condition can be stated as
\beq
\label{eq:eff_flip}
  \mu_\Gamma\equiv\frac{S}{\Gf}\gg \muQ
  \qquad ({\rm negligible~flipping}),
\eeq
or, equivalently,
\beq
    \Gf\ll\frac{\Q}{t_0}=\frac{\gamma_0}{\Q^2}.
\eeq

A convenient measure of the chiral flipping effect on the CPI, which can be used for any $\Q$ (including $\Q\ll 1$), is the dimensionless parameter
\begin{align}
\label{eq:ChiralFlipParam}
    \Gamma\equiv\frac{\Gf(1+\Q^2)}{\gamma_0}.
\end{align}
Chiral flipping has a negligible effect if $\Gamma\ll 1$. In the opposite limit of $\Gamma\gg 1$, flipping reduces the amount of magnetic helicity generated by the CPI.

Note that large $\Q\gg 1$ makes the regime of $\Gamma\gg 1$ much more likely. This regime corresponds to $\mu_\Gamma\ll \muQ$, and $\mu$ never reaches $\muQ$ since it cannot exceed $\mu_\Gamma$. Instead, the source quickly becomes balanced by chiral flipping ($S\approx \Gf\mu$ with $\mu\approx \mu_\Gamma$) while the CPI grows exponentially with the quasisteady rate  $\gamma_{\Gamma}\approx \eta \mu_{\Gamma}^2/4$. For a sufficiently long source (i.e. for sufficiently large $\Q$), the CPI eventually saturates, establishing a new steady state with $S\approx \Gf\mu+\lambda\dot H$ where a fraction $f_H=\lambda\dot{H}/S< 1$ of pumped chirality imbalance converts into magnetic helicity, while the rest $(1-f_H)=\Gf\mu/S$ is depleted by flipping. Thus, magnetic fields can be generated on some level even when $\Gamma\gg 1$. However, this requires that the CPI has time to saturate $\gamma_{\Gamma} t_0\gg 1$, which is satisfied if $\Gamma\ll \Q^{1/2}$. If $\Gamma\gg Q^{1/2}$, then the CPI growth is too slow and no magnetic field generation occurs.

In summary, the magnetic helicity generated by the CPI over the entire duration of a source with $\Q\gg 1$ is given by
\beq
\label{eq:Gamma_summary}
  H\approx\frac{\mu_0}{\lambda }\times\left\{\begin{array}{cc}
    1 &  \Gamma<1 \\
    f_H 
    & \quad 1<\Gamma\ll\Q^{1/2}\\
    0 & \Gamma\gg\Q^{1/2}\\
           \end{array}\right.
\eeq
Calculation of $f_H<1$ is complicated, because it involves the nonlinear inverse cascade of the generated helical magnetic fields during the quasisteady state $\mu\lesssim\mu_\Gamma$. This requires numerical simulations as described below.

\section{Simulations}
\label{sec:Sim}
\subsection{Setup}
\label{sec:SimSetup}

\begin{figure}
    \centering
    \includegraphics[width=\linewidth]{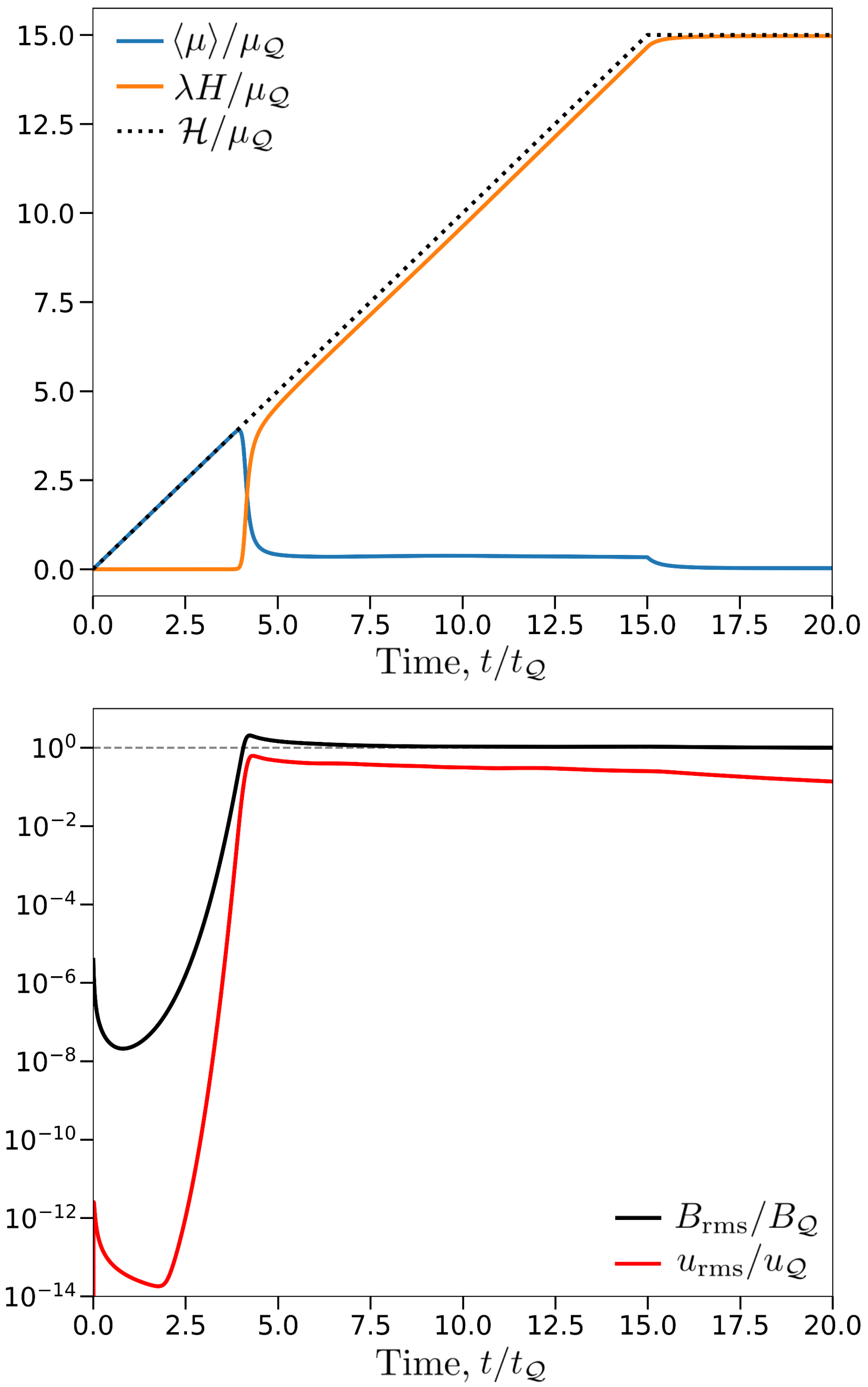}
    \caption{
    Results from a fiducial simulation with $\Q=15$ and $\Gamma=0$.
    Top: evolution of the volume-averaged chirality imbalance $\langle\mu\rangle$, magnetic helicity $H$, and total helicity $\cal H$. The source builds up  $\langle\mu\rangle\propto t$ until $t\approx 4\tQ$ when its linear growth is quenched by the saturation of the CPI. Afterwards, $\langle\mu\rangle$ relaxes to $\sim\muQ$ and the CPI generates $\lambda H$ from $\langle\mu\rangle$ with constant rate $\lambda\dot H\approx S$ while the source $S$ sustains $\langle\mu\rangle\approx0.3\muQ=const$ (until it shuts off at $t=15\tQ$). Bottom: evolution of the root-mean-square magnetic field $B_{\mathrm{rms}}$ and velocity field $u_{\mathrm{rms}}$. The horizontal dashed line indicates    $B_{\mathrm{rms}}/\BQ=u_{\mathrm{rms}}/u_{\Q}=1$, where $u_{\cal Q}=\BQ/\sqrt{4\pi\rho}$.
    }
    \label{fig:SingleRun_Q20}
\end{figure}

\begin{figure}
    \centering
    \includegraphics[width=\linewidth]{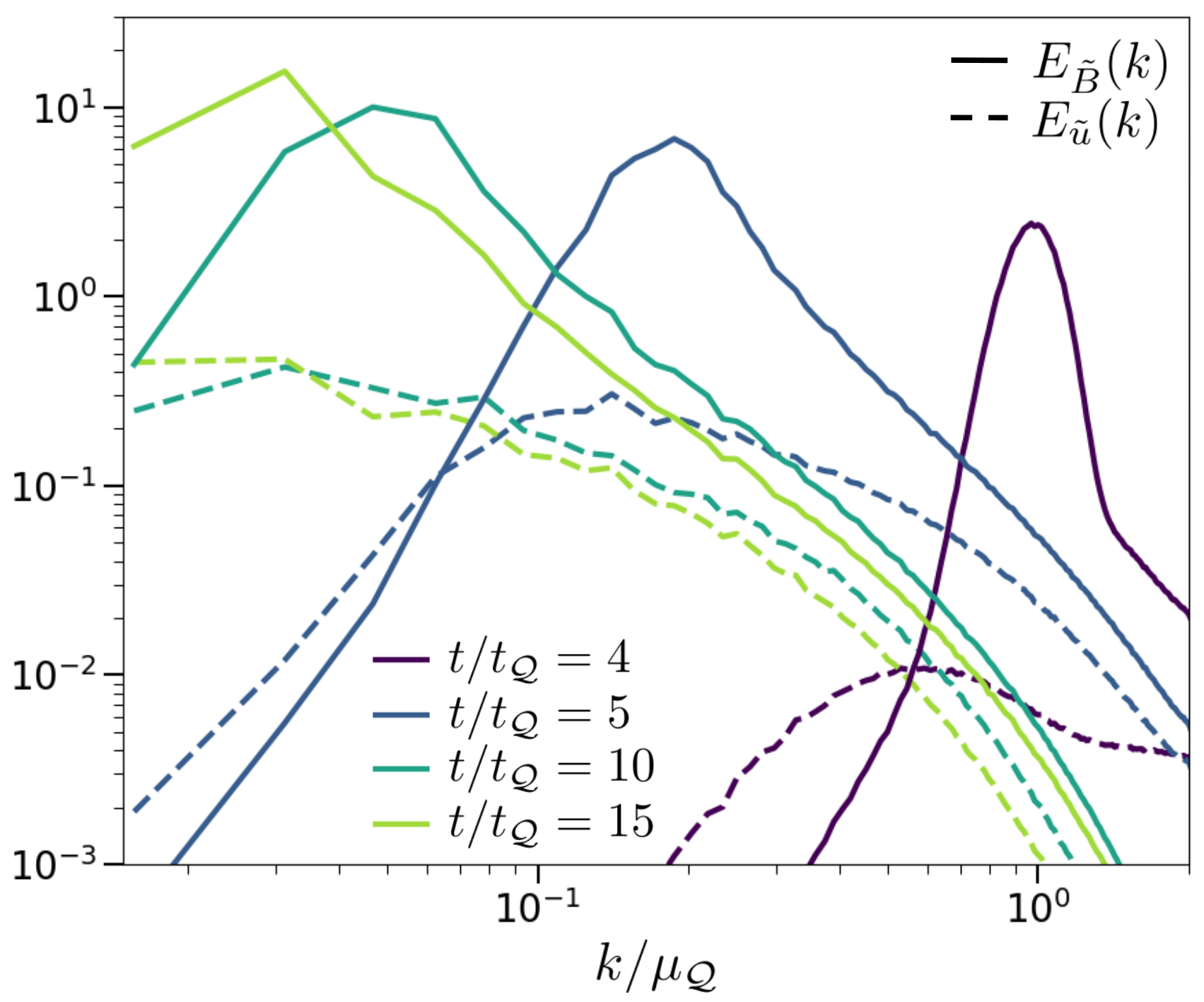}
    \caption{
    The magnetic and kinetic energy spectra, $E_{\tilde B}(k)$ (solid lines) and $E_{\tilde u}(k)$ (dashed lines), at different times (color-coded) from the simulation shown in Fig.~\ref{fig:SingleRun_Q20}.
    }
    \label{fig:EnergySpectrum}
\end{figure}
\begin{figure*}
    \centering
    \includegraphics[width=\linewidth]{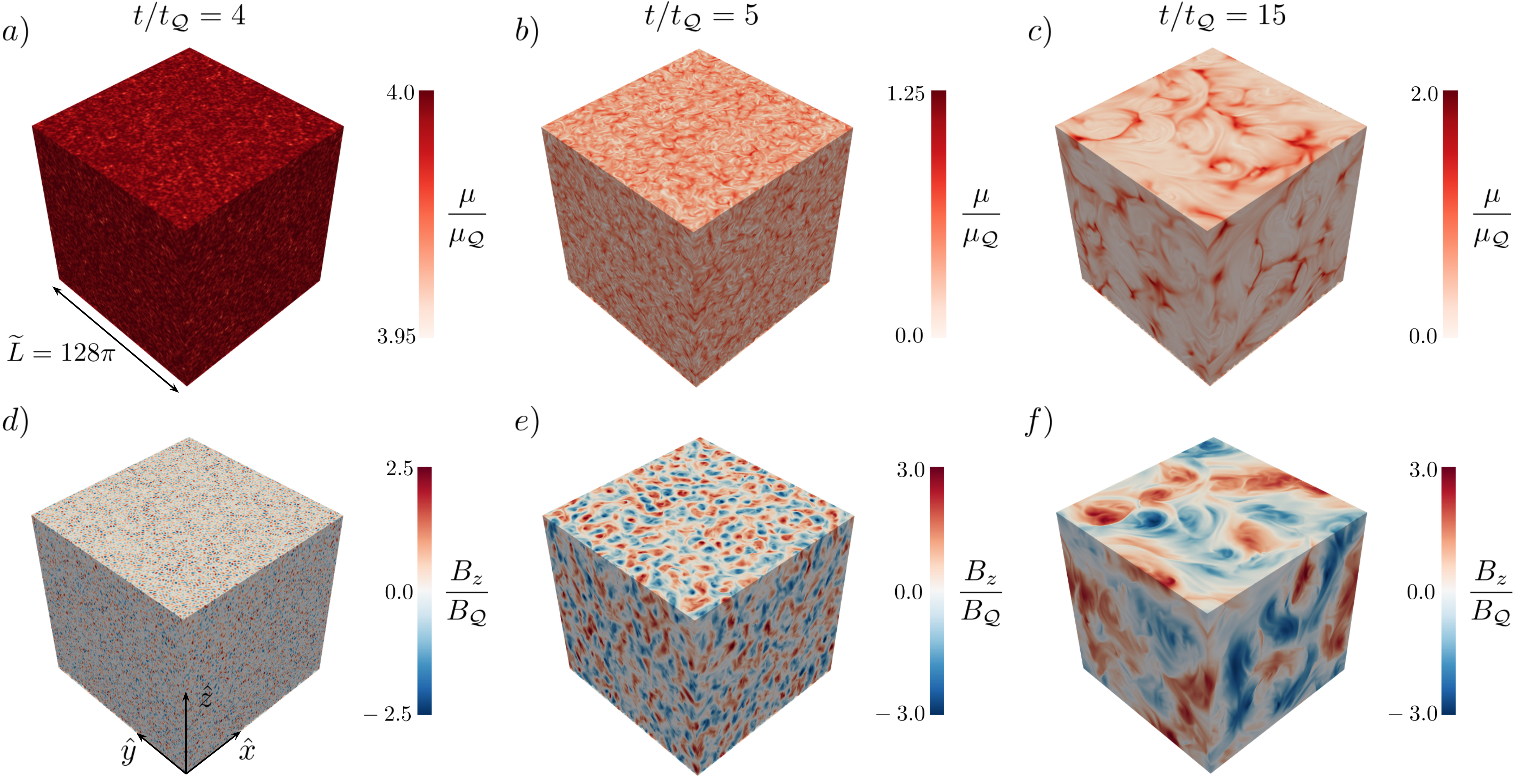}
    \caption{Visualization of the chirality imbalance $\mu$ (top row) and vertical component of the magnetic field $B_z$ (bottom row) for the fiducial simulation with $\Q=15$ analyzed in Fig.~\ref{fig:SingleRun_Q20}. The fields are shown right before CPI saturation (left column), right after the saturation (middle column), and at the end of the driving phase (right column).}
    \label{fig:3d_vis}
\end{figure*}

We now perform numerical simulations to verify the estimates  in Section~\ref{sec:Theory} and investigate the evolution of the generated magnetic fields. In the regime of $\Q\gg1$, it is convenient to define dimensionless time, wavenumbers, magnetic fields, and velocities:

\beq
  \widetilde t=\frac{t}{\tQ}, \quad
  \widetilde \mu=\frac{\mu}{\muQ}, \quad
  \widetilde\bB=\frac{\bB}{\BQ}, \quad
  \widetilde\bu=\frac{\sqrt{4\pi \rho}}{\BQ}\bu.
\eeq
The chiral MHD equations then become
\begin{align}
\label{eq:1}
    \frac{\partial\tbu}{\partial \tt} + \frac{\tbu\cdot\!\tnabla\tbu}{\chi}
    =&-\tnabla \tp+\!\frac{(\tnabla\times\tbB)\times\tbB}{\chi}+4\,{\rm Pm} \tnabla^2\tilde \bu,
    \\
\label{eq:2}
    \frac{\partial\tbB}{\partial\tt}
    =&\tnabla\times\left(\frac{\tbu\times\tbB}{\chi} +4\, \tmu\tbB \right)+ 4\,\tnabla^2\tbB,
    \\
\label{eq:3}
    \frac{\partial\tmu}{\partial \tt} + \frac{\tbu\cdot\!\tnabla\tmu}{\chi}
    =&\,8\left(\tbB\cdot(\tnabla\times \tbB)-\tmu\tbB^2\right)\nonumber\\&+\tS-
    \frac{\Gamma\tmu}{1+\Q^{-2}},
    \\
\label{eq:4}
    \tnabla\cdot\tbu
    & =0,
\end{align}
where $\tp=\tQ (4\pi\lambda /\rho)^{1/2}p$, $\tnabla=\nabla/\muQ$, and
\beq
  \chi=\frac{1}{2}\eta(\pi\rho\lambda)^{1/2}, \qquad \Pm=\frac{\nu}{\eta}.
\eeq
The dimensionless source is
\beq
  \tS=1-\Theta(\tilde t-\Q),
\eeq
where $\Theta$ is the Heaviside step function. The source pumps the dimensionless total helicity to $\tilde{\cal H}=\Q$ over the time interval $0\leq\tilde t\leq{\cal Q}$. 

\Eqs~(\ref{eq:1})-(\ref{eq:4}) determine the evolution of $\tmu$, $\tbu$, and $\tbB$ and we use initial conditions $\tmu=0$, $\tbu=0$, and $\tbB=\tbB_0$ (a small initial magnetic field to seed the CPI). The problem has four dimensionless parameters: $\Q$, $\Gamma$, $\chi$, and $\Pm$. All models presented below have $\chi=0.02$, and $\Pm=1$. The seed fields are initialized from a Gaussian random magnetic vector potential that results in a magnetic energy spectrum $E_{\tB}(\tk)\approx 10^{-14}\tk^4$ with $|\tbB_0|\sim 10^{-6}$.

We note that $\chi$ is related to the magnetic Reynolds number $\Rm\sim|\tnabla\times( \tbu\times\tbB)|/4\chi|\tnabla^2\tbB|$. When $\Rm\gg 1$, magnetic fields cannot quickly diffuse through the fluid, so they efficiently excite fluid motions through the Lorentz force. The relevant value of $\Rm$ may be evaluated with $\tnabla=i\tbk$ using the characteristic CPI wavenumber $\tk\sim\tmu$, which gives $\Rm\gg 1$ for both early Universe and PNS \citep{schober2018laminar}. We have found that the helicity evolution in the quenched (turbulent) phase of the simulations is nearly independent of $\chi$ when $\Rm\gg1$. The selection of the fiducial value of $\chi=0.02$ was justified by a parameter scan of $\chi$ (Appendix~\ref{ap:Validation}). 

The chiral MHD equations are solved using the Dedalus pseudo-spectral code \citep{burns2020dedalus}. The domain is a cube with triply periodic boundary conditions, a resolution of $N^3=256^3$ Fourier modes, and side length $2\pi N/4\muQ$. The box size is chosen to capture the full range of scales during the evolution. The simulation must resolve both the quenching phase when fields with the largest wavenumbers $\tk\sim\tmu= {\cal O}(1)$ at $\tilde t={\cal O}(1)$ are amplified by the CPI and the subsequent steady state of magnetic helicity growth when the wavenumbers of the fields decrease down to $\tk={\cal O}(Q^{-1})$ near $\tilde t=\Q$.

\subsection{Evolution without chiral flipping}
\label{sec:SimSource}

We first present a fiducial simulation with $\Q=15$ and no chiral flipping $\Gamma=0$. The final magnetic helicity $H$ is expected to approach $\Htot/\lambda$ (Section~\ref{sec:TheorySource}), where $\Htot=\Q\muQ$ ($\tHtot=\Q)$. A parameter scan of $\Q$ (Appendix~\ref{ap:Validation}) confirms that $\Q=15$ is sufficiently large to exhibit the quenching dynamics expected in the limit of $\Q\gg1$.

The results of the simulation are shown in Fig.~\ref{fig:SingleRun_Q20}. One can see three evolution phases: 
(1) The total helicity is mostly stored in the pumped chirality imbalance $\mu=St$ of the particles while the CPI amplifies the seed fields. The conversion rate of $\mu$ into $H$ by the CPI is negligible. This initial phase lasts until $\tt\approx 4$. 
(2) At $\tt>4$, the CPI now converts $\mu$ into $H$ at the same rate as $\mu$ is pumped by the source. As a result, the magnetic helicity grows  $\lambda H=St-0.3\muQ$  while the chirality imbalance remains constant $\mu\approx0.3\muQ$. This steady state lasts until the source stops at $\tt=15$.
(3) At $\tt>15$, $\mu$ decays and the final magnetic helicity levels out at $\lambda H\approx\Q\muQ=15\muQ$.

Note that the CPI becomes dynamically significant at $\tt\approx 4$, later than $\tt\sim 1$ predicted by the simple estimate in Section~\ref{sec:TheorySource}. This delay (and overshoot of $\mu>\muQ$) is due to the finite amount of time needed to amplify the seed fields, which was neglected in Section~\ref{sec:TheorySource}. Field amplification during the linear growth of $\tmu=\tt$ can be described by the magnetic energy spectrum generated by the CPI,
\beq   
E_{\tB}(\tk,\tt)\approx E_{\tB}(\tk,0)\exp\left[\int_{\tk}^{\tt}2\tgamma(\tk,\tt') \,d\tt'\right],
\eeq
where $E_{\tB}(\tk,0)$ is the initial energy spectrum, and $\tgamma(\tk,\tt)=4\tk(\tmu-\tk)\approx 4\tk(\tt-\tk)$ is the dimensionless growth rate of a mode with wavenumber $\tk$ at time $\tt$. The energy spectrum $E_{\tB}(\tk,\tt)$ sharply peaks at a moving wavenumber $\tk_{\rm pk}(\tt)=\tt/3$. The maximally helical field is thus concentrated at a length scale $\sim 1/\tk_{\rm pk}$ and has magnetic helicity $\tH\approx \tB_{\rm rms}^2/\tk_{\rm pk}\approx E_{\tB}(\tk_{\rm pk},\tt)$. Growth of $\tmu$ ends once a large fraction of $\tmu$ is converted into magnetic helicity $\tH\sim\tmu$. This occurs at time $\tt_{\rm sat}$ that satisfies

\begin{align}
\label{eq:T_quench_numerical}
    \tt_{\rm sat}=\left[\frac{27}{16}\ln\left(\frac{\tt_{\rm sat}}{E_{\tB}(\tt_{\rm sat}/3,0)}\right)\right]^{1/3}.
\end{align}
Using $E_{\tB}(\tk,0)\sim 10^{-14}$, \Eq~(\ref{eq:T_quench_numerical}) predicts $\tilde t_{\rm sat}\approx 3.8$, in agreement with the saturation time observed in Fig.~\ref{fig:SingleRun_Q20}.

After saturation $\tt>\tt_{\rm sat}$, the spectrum widens and moves to progressively larger scales $\tL=\tk^{-1}$ (Fig.~\ref{fig:EnergySpectrum}) while the magnetic helicity continues to grow as $\tH=\tt-0.3$ (in dimensionless form). Remarkably, the magnetic helicity $\tH\sim \tL\tB_{\rm rms}^2$ grows by increasing the length scale $\tilde L$, not by increasing the field strength, as $\tB_{\rm rms}\sim 1$ stays constant (Fig.~\ref{fig:SingleRun_Q20}). Note also that $\tilde u_{\rm rms}\sim\tilde 0.2 B_{\rm rms}$ at $\tt>\tt_{\rm sat}$. Hence, the magnetic Reynolds number grows as $\Rm\sim 0.2\tB_{\rm rms} \tL/4\chi\propto \tt$. It reaches a maximum of $\Rm\sim0.05\Q/\chi\approx40$ at the end of the driving phase near $\tilde t\sim 15$ when $\tilde L\sim\Q= 15$.

The qualitative properties of the evolution discussed above are also visible in the spatial plots shown in Fig.~\ref{fig:3d_vis}. One can see that at $\tilde t=4$, right before the linear growth of $\tmu=\tt$ is quenched, the chirality imbalance $\tilde\mu=4$ is approximately uniform across the box, and the generated magnetic fields are concentrated at small scales corresponding to $\tk\sim 1$. By $\tt=5$, the average value of $\tmu$ has relaxed to $\langle\tmu\rangle\approx 0.3$, and significant fluctuations of $\tmu$ have developed. The subsequent evolution proceeds with the constant $\langle\tmu\rangle\approx 0.3$ and $\tB_{\rm rms}\approx 1$ while the spatial scale of the fluctuations grows with time, approaching the box size at $\tt\sim 15$
\footnote{The large spatial scale of magnetic fields that develop by the end of the driving phase, $ \tilde L\sim \tilde H/\tilde B_{\rm rms}^2\approx\Q$, requires a large simulation box with the required resolution scaling as $N^3\propto \Q^3$.}.
After the source stops at $\tt=15$, the usual inverse cascade with constant helicity $H$ will occur with $\tB_{\rm rms}\propto \widetilde L^{-1/2}$ at $\tt\gg 15$ \citep{woltjer1958theorem,taylor1986relaxation,biskamp1999decay,hirono2015self,brandenburg2017turbulent,hosking2021reconnection}; however, this is not followed by our simulation, which stops at $\tt=20$.

\subsection{Evolution with chiral flipping}
\label{sec:SimFlipping}

\begin{figure}
    \centering
    \includegraphics[width=\linewidth]{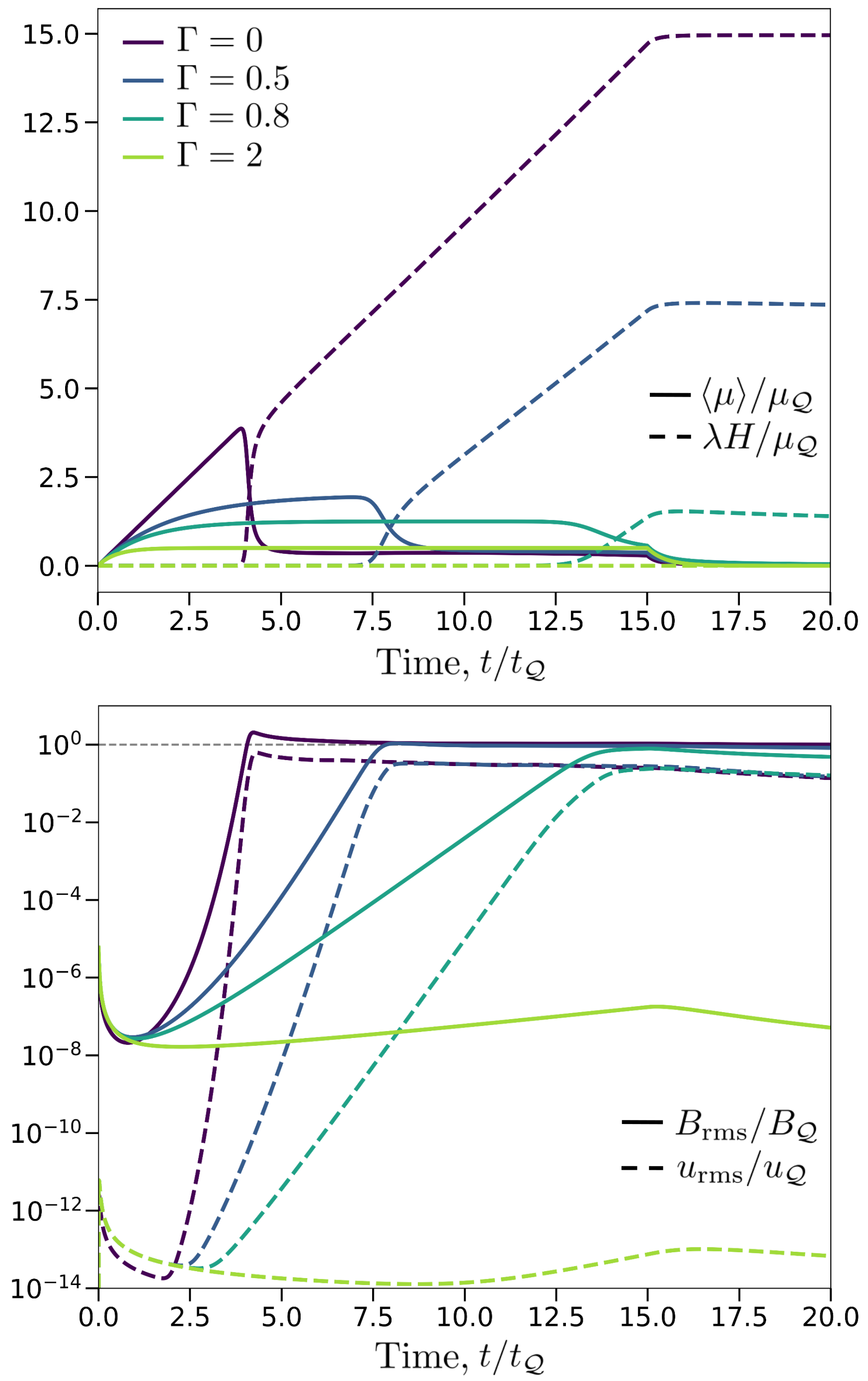}
    \caption{
    A set of simulations varying the chiral flipping parameter $\Gamma$ (color-coded) with fixed $\Q=15$. The $\Gamma=0$ case is the same simulation as analyzed in Fig.~\ref{fig:SingleRun_Q20}. 
    Top: evolution of the volume-averaged chirality imbalance $\langle\mu\rangle$ (solid lines) and the magnetic helicity $H$ (dashed lines).
    Magnetic helicity generation by the CPI is suppressed by flipping once  $\Gamma$ becomes larger than ${\cal O}(1)$. 
    Bottom: same quantities as in Fig.~\ref{fig:SingleRun_Q20}. 
    }
    \label{fig:Gamma_scan}
\end{figure}

\begin{figure}
    \centering
    \includegraphics[width=\linewidth]{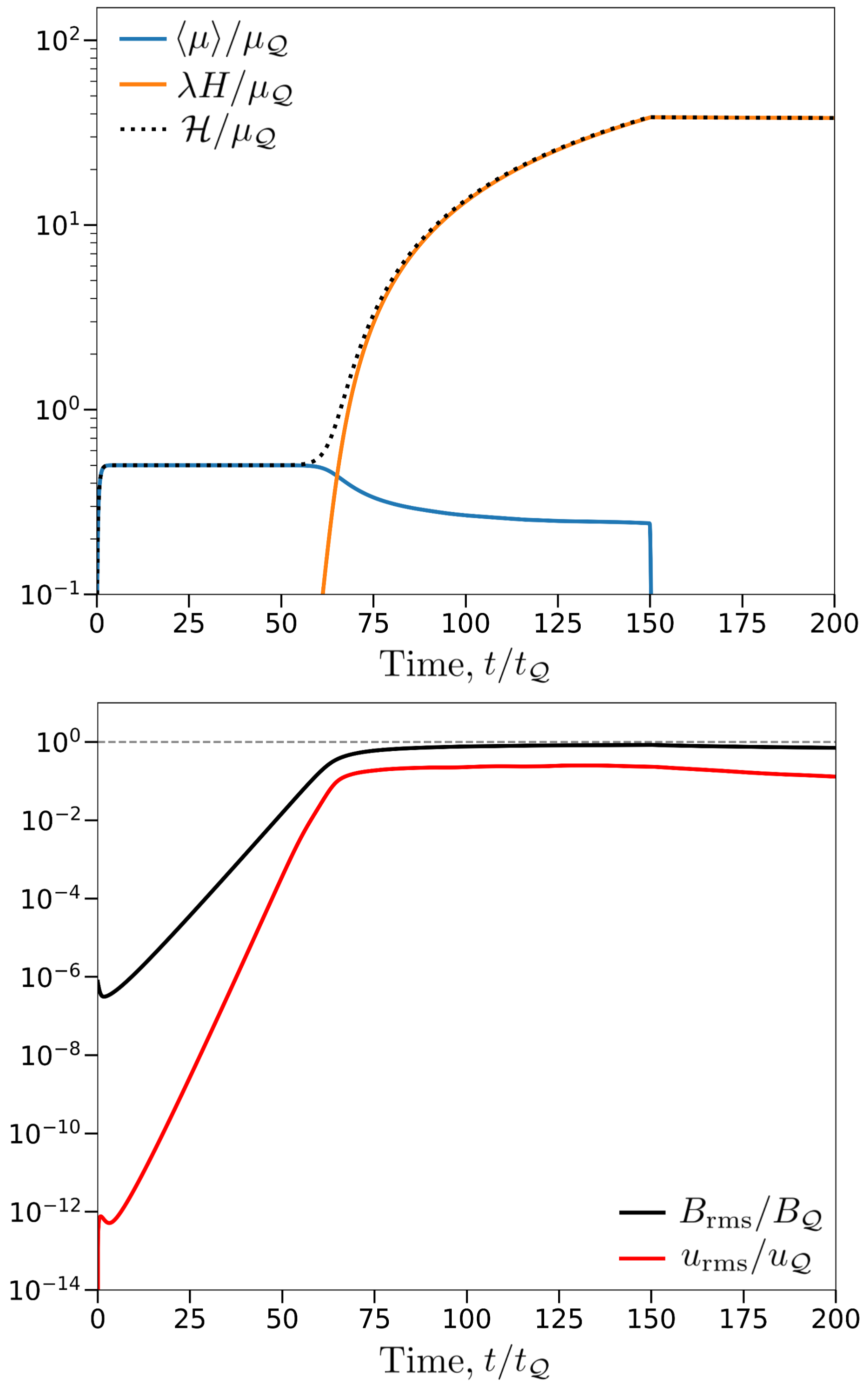}
    \caption{
    A simulation with $\Gamma=2,$ $\Q=150$ in the flipping-limited regime $1<\Gamma\ll Q^{1/2}$ where the CPI has enough time to saturate before the source shuts off $\tt_{\Gamma,\rm sat}<\Q$.
    Top and bottom panels show the same quantities as in Fig.~\ref{fig:SingleRun_Q20}.
    When the CPI saturates at $\tt_{\Gamma,\rm sat}\approx 65$, the system switches from a steady state limited by flipping $\tilde S\approx\Gamma \tilde \mu_\Gamma$ to one of magnetic helicity generation $\dot \tH=\tS-\Gamma \tmu\approx const$ (though less efficiently than when $\Gamma=0$).  
    }
    \label{fig:Gamma2_Q150}
\end{figure}

We now introduce chiral flipping $\Gamma\neq0$ and examine its impact on CPI growth and saturation. Below, we compare simulations (Fig.~\ref{fig:Gamma_scan}) with increasing $\Gamma=0.5$, $0.8$, $2$ against the model without flipping using our fiducial setup with $\Q=15$. We then examine one additional simulation with $\Gamma=2$ and $\Q=150$ (Fig.~\ref{fig:Gamma2_Q150}).

Chiral flipping imposes the ceiling $\tilde \mu\leq\tilde \mu_\Gamma\approx\Gamma^{-1}$ (Section~\ref{sec:TheoryFlipping}). As $\Gamma$ is increased from zero, it will first affect the transient overshoot of $\tmu$ at $\tt\approx \ttsat$, when the largest $\tmu\approx\ttsat$ is reached (Section~\ref{sec:SimSource}). Thus, the overshoot is reduced when $\Gamma>\ttsat^{-1}$, i.e. $\Gamma>0.25$ for the parameters of our simulations. Notably, when $\ttsat^{-1}<\Gamma<1$, flipping affects only the initial overshoot but not the later steady state that is established at $\tt>\ttsat$. This is illustrated by the simulation with $\Gamma=0.5$: $\tmu$ remains below $\tmu_\Gamma\approx 2$ during the initial transient phase, causing the CPI to develop somewhat slower than in the case of $\Gamma=0$, but the steady state $\lambda \dot H\approx S$ is eventually established (by $\tt\approx 8$).

The steady state is altered when $\Gamma>1$, as flipping quickly enforces $\tmu\approx\tmu_\Gamma=\Gamma^{-1}<1$ (for $\tt>\Gamma^{-1}$). The initial amplification phase of the seed fields then occurs with constant $\tmu\approx \tmu_\Gamma$ and the CPI develops at $\tk\approx \tk_{\rm pk}=\tmu_\Gamma/2$ with growth rate $\tgamma=\Gamma^{-2}$. The time $\tt_{\rm sat,\Gamma}$ needed for the fields to saturate may be estimated similarly to that of $\ttsat$ in Section~\ref{sec:SimSource}. Using the magnetic energy spectrum $E_{\tB}(\tk,\tt)\approx E_{\tB}(\tk,0)\exp(2\tgamma\tt)$ (which has a narrow peak at $\tk\approx \tk_{\rm pk}$) and magnetic helicity $\tH\approx E_{\tB}(\tk_{\rm pk},\tt)$, the time $\tt_{\rm sat,\Gamma}$ is found from the condition $\tH \sim \tmu_\Gamma$, which gives 

\begin{align}
\label{eq:Updatedtsat}
    \tt_{\rm sat,\Gamma}\approx  0.5\Gamma^2\left|\ln\left[\Gamma E_{\tB}(\tk_{\rm pk},0)\right]\right|,\qquad \Gamma>\ttsat^{-1}.
\end{align}
Recall that the source  operates during a limited time interval $0<\tt<\Q$. If $\tt_{\rm sat,\Gamma}>\Q$, then $\tH$ will not even reach  $\tmu_\Gamma$, and magnetic field generation by the CPI becomes an exponentially small effect. This occurs when
\begin{align}
\label{eq:UpdatedGamma}
    \Gamma>\Gamma_{\max}=\frac{\Q^{1/2}}{\left|0.5\ln[\Gamma E_{\tB}(\tk_{\rm pk},0)]\right|^{1/2}} \qquad (\rm no\;CPI).
\end{align}

This is a tighter bound than~\Eq~(\ref{eq:Gamma_summary}), accounting for the large number of $e$-foldings needed for saturation. Equation~\ref{eq:UpdatedGamma} predicts $\Gamma_{\max}\approx1$ for our simulation parameters. This is illustrated by the $\Gamma=0.8$ simulation where the CPI saturates at $\tilde t\approx \Q$ right as the source stops. Since $\Gamma_{\max}\approx1$ in our setup with $\Q=15$, CPI saturation does not occur for $\Gamma>1$, as confirmed by the $\Gamma=2$ simulation where the CPI is  completely inactive.

To reach the regime $1<\Gamma<\Gamma_{\max}$, we perform an additional simulation with $\Gamma=2$ and a much larger $\Q=150$ such that $\Gamma_{\max}\approx 3$. The only modification to the simulation setup was increasing the box size by a factor of $8$, as CPI wavenumbers remain small $\tk\sim\tmu\leq\Gamma^{-1}$. The result is presented in Fig.~\ref{fig:Gamma2_Q150}. Initially, $\tmu=\tmu_\Gamma=0.5$ while the CPI grows. Then, the CPI saturates at $\tt_{\rm sat,\Gamma}\approx65$ before the source shuts off at $\tt=150$. This slightly reduces the imbalance to $\tmu=(1-f_H)\tmu_{\Gamma}$ with a measured $f_H\approx 0.5$, allowing the magnetic helicity  to then grow at a constant rate $\dot \tH\approx f_H= 0.5$. When the source shuts off, the magnetic helicity reaches $\tH\approx\Q f_H(1-\tt_{\rm sat,\Gamma}/\Q)\approx 40$, below the $\tH\approx\Q=150$ that would have been generated without flipping.

\section{Astrophysical applications}
\label{sec:Apps}
\begin{figure}
    \centering
    \includegraphics[width=\linewidth]{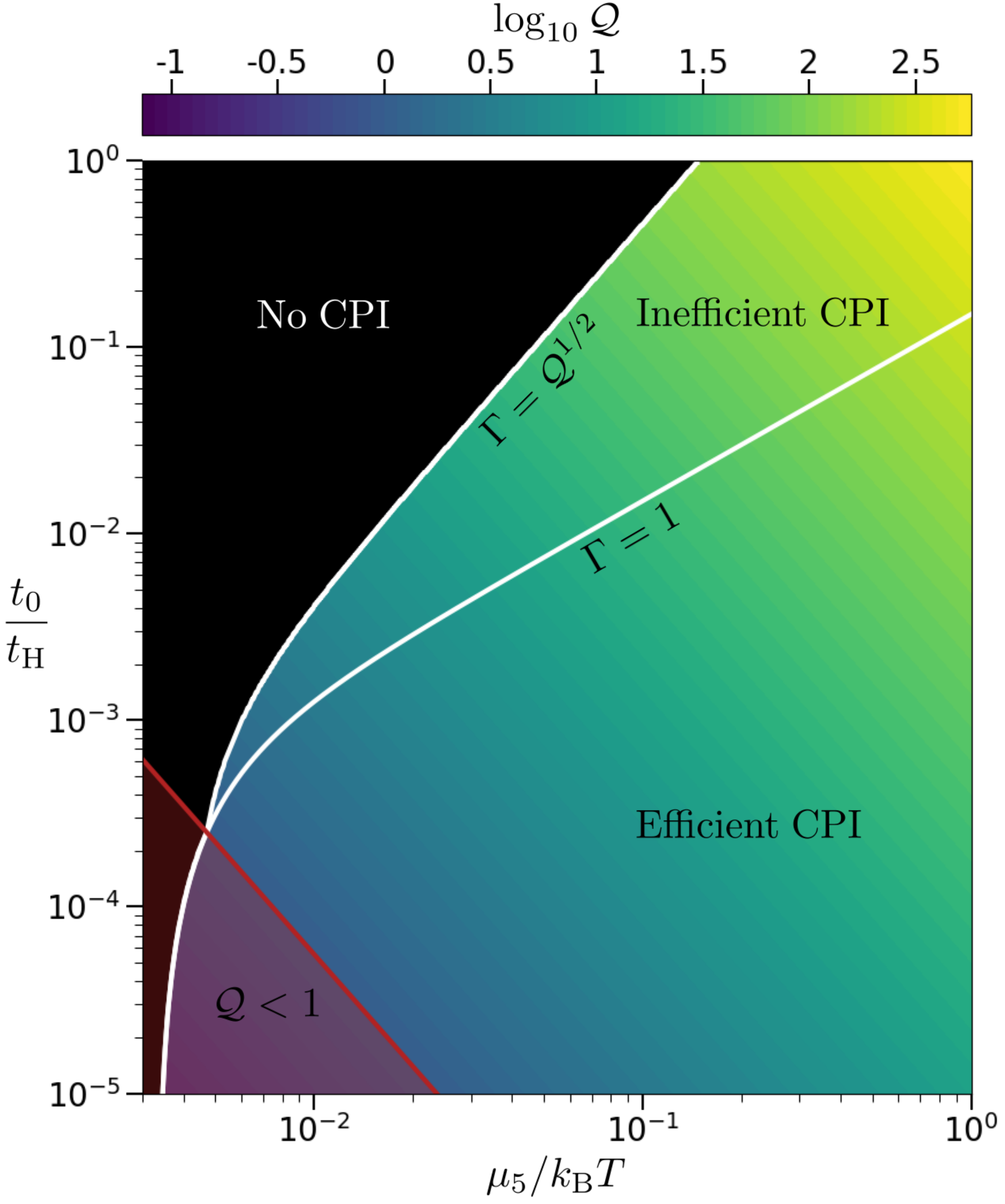}
    \caption{
    Regimes of a possible chiral dynamo in the early Universe near the electroweak transition ($k_{\rm B}T=100$\,GeV), depending on the strength $\mu_0$ and duration $t_0$ of the chiral source $S=\mu_0/t_0$. 
    Color indicates the values of $\Q$.
    The reduction of the CPI rate due to $\Q>1$ occurs in the broad parameter space above the red region. White curves indicate contours of the chiral flipping parameter $\Gamma$ defined in \Eq~(\ref{eq:ChiralFlipParam}). 
    The plot has three zones that approximately represent three different regimes of the chiral dynamo. 
    (1) Chiral flipping weakly affects the CPI where $\Gamma<1$; most of the chirality imbalance $\mu_0$ eventually converts into magnetic helicity. 
    (2) The CPI still operates where $1<\Gamma<\max\{1,\Q^{1/2}\}$, but chiral flipping is significant and erases a fraction of $\mu_0$ during the chiral dynamo, reducing its efficiency.
    (3) The CPI is exponentially suppressed where $\Gamma>\max\{1,\Q^{1/2}\}$ (black region). More accurate boundaries between these regimes are shifted down and to the right by a logarithmic factor, which depends on the seed amplitude of the magnetic fields (see Section~\ref{sec:SimFlipping}).}
    \label{fig:Early_universe}
\end{figure}
\subsection{Early Universe}

Calculations of the chiral flipping rate in the early Universe \citep{campbell1992baryon,joyce1997primordial,boyarsky2021evolution} may be summarized as follows
\beq
   \Gf\approx 2\cdot10^{13}\left(T_{100}+T_{100}^{-1}\right) {\rm s}^{-1},
\eeq
where $T_{100}\equiv k_{\mathrm{B}}T/100\,$GeV.
Below we focus on the moment right after the electroweak transition at $k_{\rm B}T\approx 100$\,GeV. Note that  at this moment the flipping rate has its lowest value $\Gf\approx 4\times 10^{13}$\,s$^{-1}$.

The CPI model assumes that the injected chirality imbalance $\mu_5=\hbar c\mu_0/4\alpha$ is below $k_{\rm B}T$, and the characteristic injection timescale $t_0$ does not exceed the Hubble timescale $t_{\rm H}\approx 5\cdot10^{-11}T_{100}^{-2}\,$s. The CPI has a chance to develop if its maximum possible growth rate $\gamma_0=\eta\mu_0^2/4$ exceeds the flipping rate $\Gf$. Combining these conditions, we conclude that a successful CPI at $k_{\rm B}T\sim 100$\,GeV should be looked for in the following parameter space:
\begin{align}
\label{eq:sourcerange}
    3\cdot10^{-3} <\frac{\mu_5}{k_{\rm B}T}<1, \qquad t_0\leq t_{\rm H}.
\end{align}
Here, we used the magnetic diffusivity $\eta\approx 7\cdot 10^{-10}\,T_{100}^{-1}\,\mathrm{cm}^2\mathrm{s}^{-1}$ \citep{arnold2000transport}. 

The CPI is controlled by the parameters $\Q$ and $\Gamma$ defined in \Eqs~(\ref{eq:Q}) and (\ref{eq:ChiralFlipParam}). The dependence of $\Q$ and $\Gamma$ on the source parameters $\mu_5/k_{\rm B}T$ and $t_0/t_{\rm H}$ is presented in Fig.~\ref{fig:Early_universe}. ``Fast" sources ($\Q<1$) build up $\mu_0$ and lead to a CPI with the maximum growth rate $\gamma_0$. This regime occurs in a small region of the parameter space where $t_0<6\cdot 10^{-4}t_{\rm H}$ and $\mu_5/k_{\rm B}T<7\cdot 10^{-3}(t_0/10^{-4}t_{\rm H})^{-1/2}$. This region mostly has $\Gamma<1$, so the CPI is weakly affected by flipping.

More realistic sources operate on a longer timescale $t_0\lesssim t_{\rm H}$ with $\Q\gg 1$. Such ``slow'' sources occupy most of the parameter space in Fig.~\ref{fig:Early_universe}. The effect of chiral flipping on the CPI is controlled by $\Gamma$ as summarized in \Eq~(\ref{eq:Gamma_summary}); the corresponding zones of the parameter space are indicated in Fig.~\ref{fig:Early_universe}. In particular, consider a CPI model with $t_0\approx t_{\rm H}$. One can see that flipping affects the CPI ($\Gamma>1$) for all $\mu_5<k_{\rm B}T$. The CPI with $t_0\approx t_{\rm H}$ is completely suppressed ($\Gamma>\Q^{1/2}$) if $\mu_5<0.1 k_{\rm B}T$. Flipping becomes negligible ($\Gamma< 1$) only for faster sources with $t_0\lesssim 0.1t_{\rm H}$ and $\mu_5/k_{\rm B}T>t_0/0.1t_{\rm H}$. 

\subsection{Protoneutron stars}

We first emphasize that even an efficient chiral dynamo cannot generate global, magnetar-level magnetic fields due to the limited helicity budget. The budget is bounded above by the chirality imbalance $\mu_5\lesssim\mu_e/2\sim100\,$MeV available after capture of all left-handed electrons \citep{ohnishi2014magnetars}, which gives
\begin{align}
    H<\frac{2\alpha\mu_e}{\hbar c\lambda}\sim 4\cdot 10^{24}\,\mathrm G^2\,\mathrm{cm}\left(\frac{\mu_e}{200\,\rm MeV}\right)^3.
\end{align}
Inverse cascade of such maximally helical magnetic fields  (conserving $H\sim LB^2$) to length scales comparable to the PNS radius $L\sim 10^6\,\rm cm$ would result in fields $B\sim(H/L)^{0.5}\sim2\cdot10^9\,$G that are far weaker than the magnetar fields $B\gtrsim 10^{15}$\,G \citep{kaspi2017magnetars} \footnote{Previous studies have suggested that magnetar-strength fields $B\sim2\alpha\mu_e/\hbar c\sqrt{\lambda}\lesssim 10^{18}\,$G could be generated if the chirality imbalance reaches $\mu_5\sim\mu_e/2$ before the CPI is triggered. However, such fields would be generated on microscopic scales $L\sim2\pi\hbar c/\alpha\mu_e\lesssim10^{-10}\,$cm and their inverse cascade would still lead to similar $B\sim10^{9}\,$G fields on global scales.}. Such moderate fields would be comparable to the flux-frozen fields $10^8-10^{12}\,$G likely inherited from progenitor cores \citep{spruit2008origin,ferrario2015magnetic}.

However, it is easy to see that chiral flipping completely suppresses the CPI in PNS. Chirality imbalance is sourced in the PNS at the depletion rate of the electron population \citep{grabowska2015role,sigl2016chiral}. This could drive the  chiral chemical potential to a maximum $\mu_5\lesssim \mu_e/2$ over a timescale $t_0\sim1\,$s. The magnetic diffusivity in the PNS interior is controlled by scattering of the degenerate electrons by protons and is given by \citep{thompson1993neutron}
\begin{align}
\eta\approx  3\cdot 10^{-5}
\left(\frac{\rho}{10^{14}\,\mathrm{g}\,\mathrm{cm}^{-3}}\right)^{-1/3} \mathrm{cm}^2\,\mathrm{s}^{-1}.
\end{align}
Combining these characteristic values of $\mu_5$, $t_0$, and $\eta$, we find the quenching parameter  
\begin{align}
\label{eq:Q_PNS}
    \Q\sim 5\cdot 10^5\left(\frac{\mu_e}{200\,\mathrm{MeV}}\right)^{2/3}\left(\frac{t_0}{1\,\mathrm{s}}\right)^{1/3}.
\end{align}
Note that, without flipping, the CPI would quench the chiral chemical potential at $\mu_5\sim \mu_e/2\Q\sim 200\,\mathrm{eV}$, well below the thermal energy scale  $ k_{\rm B} T\sim30\,\mathrm{MeV}$.

The flipping rate in a PNS plasma is dominated by Rutherford scattering and given by $\Gf\sim10^{14}\,\mathrm{s}^{-1}(200\,\mathrm{MeV}/\mu_e)$ in the appropriate limit $\mu_5\ll k_{\rm B} T$  \citep{grabowska2015role,sen2026chiral}. Then, using the estimate for $\Q$ in \Eq~(\ref{eq:Q_PNS}), one finds 
\begin{align}
    \Gamma\approx\frac{\Gf \Q^2}{\gamma_0}\sim 2\cdot 10^{8}\left(\frac{\mu_e}{200\,\mathrm{MeV}}\right)^{-5/3}\left(\frac{t_0}{1\,\mathrm{s}}\right)^{2/3}.
\end{align}
One can see that $\Gamma\gg\Q^{1/2}$, and hence the CPI is completely (exponentially) suppressed (Equation~\ref{eq:Gamma_summary}). Thus, the chiral helicity budget is entirely erased by chiral flipping without generating magnetic fields.

\section{Conclusions}

We have investigated chiral dynamos with a chirality imbalance $\mu_5$ pumped by a source on a finite timescale $t_0$. We have found that the CPI develops with a limited rate $\gamma\approx\gamma_0/(1+\Q^2)$, where $\Q= (\gamma_0 t_0)^{1/3}$ and $\gamma_0$ corresponds to models with $t_0=0$ (\Eq~\ref{eq:gamma}). The slowdown of CPI by the factor of $1+\Q^2$ has important consequences when the chiral dynamo competes with the deteriorating effect of chiral flipping $\Gf$. In the case of $\Q<1$, chiral flipping  suppresses the CPI if $\Gf>\gamma_0$. In the more realistic case of $\Q\gg 1$, flipping reduces the final magnetic fields generated by the CPI if $\Gf>\gamma_0/\Q^2$ and completely suppresses them if $\Gf>\gamma_0/\Q^{3/2}$. We conclude that the approximate general condition for the complete suppression of the CPI can be stated as $\Gf>\gamma_0/(1+\Q^{3/2})$, which can be used at any $\Q$.

Both examples of astrophysical chiral plasmas discussed in this paper have $\Q\gg 1$. In PNS, $\Q={\cal O} (10^5)$ and $\Gf\gg \gamma_0/\Q^{3/2}$, which implies the complete suppression of chiral dynamo. For the possible CPI near the electroweak phase transition in the early Universe, we estimate $\Q={\cal O} (10^2)$ unless $t_0$ is very much smaller than the Hubble time $t_{\rm H}$. We find that the strong suppression of the CPI may be marginally avoided in scenarios of a source that tries to pump $\mu_5\sim k_{\rm B}T$ on a timescale $t_0\lesssim t_{\rm H}$ (Fig.~\ref{fig:Early_universe}).

\begin{acknowledgments}
We thank Michael Landry and Nitya Nigam for initial discussions. This work is supported by NSF Grant No. AST-2408199. A. M. B. also acknowledges support by NASA ATP Grant No. 80NSSC24K1229 and Simons Foundation Award No. 446228.
\end{acknowledgments}

\vspace{5mm}

\appendix

\section{Validation of the fiducial parameters}
\label{ap:Validation}

\begin{figure}
    \centering
    \includegraphics[width=\linewidth]{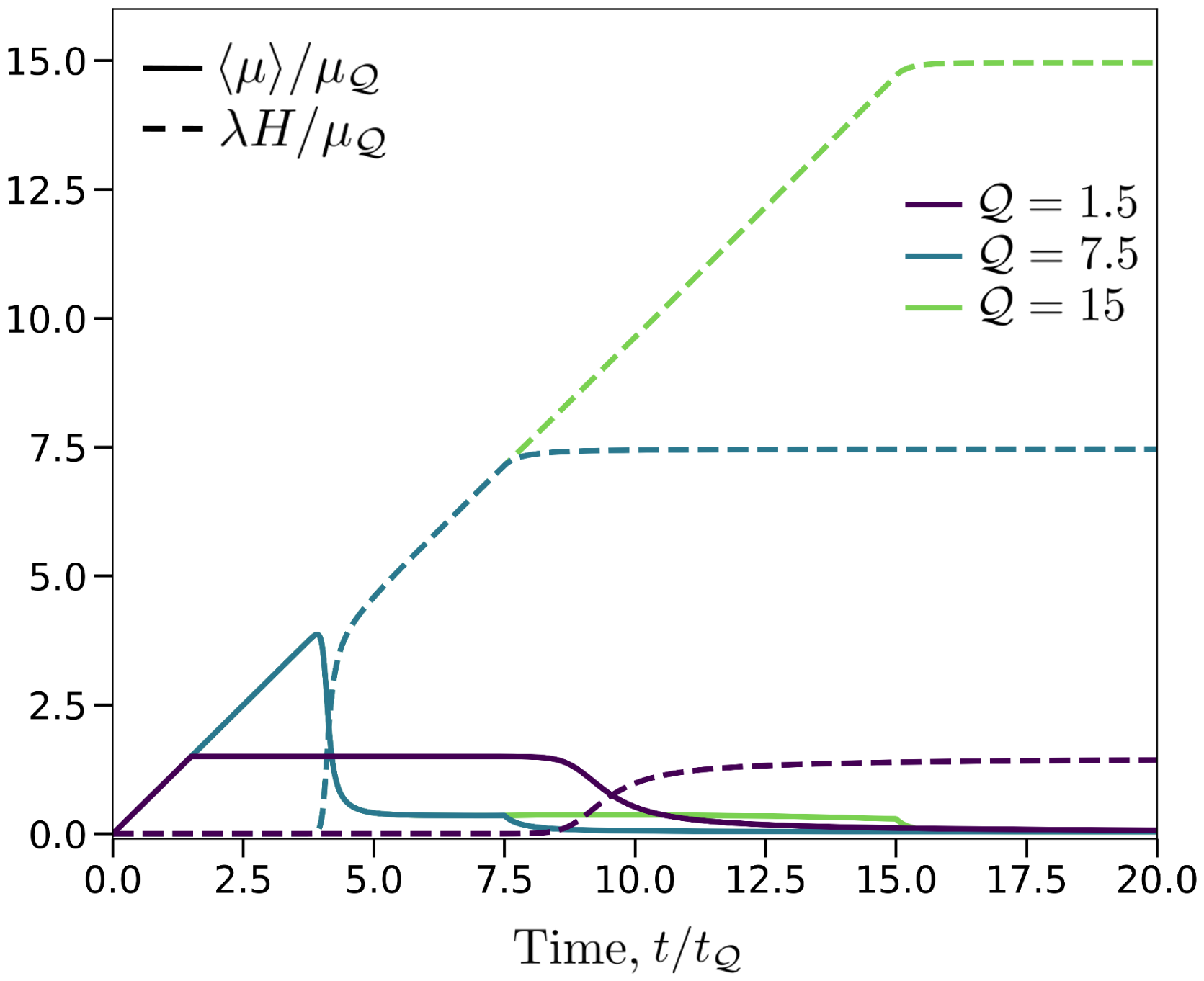}
    \caption{
    Comparison of simulations with $\Q=1.5,7.5,$ and 15. All three simulations have $\chi^{-1}=50$ and no chiral flipping. The source acts during the time interval $0\leq t/\tQ\leq\Q$. In both simulations with large $\Q$ ($7.5$ and 15), the growth of the chirality imbalance $\langle\mu\rangle$ (solid curves) is quenched once the CPI saturates (near $t/\tQ\sim 4$), resulting in subsequent linear growth of the magnetic helicity $H$ (dashed curves). By contrast, in the $\Q=1.5$ simulation, $\langle\mu\rangle$ quickly reaches its maximum and the source stops before the CPI had a chance to saturate; the CPI saturates at later times $t\gtrsim 9\tQ$.
    }
    \label{fig:Q_scan}
\end{figure}

For the simulations presented in Section~\ref{sec:Sim} we chose $\Q=15$ and $\chi=0.02$ as fiducial parameters. Here we present additional simulations with different $\Q$ and $\chi$.

Fig.~\ref{fig:Q_scan} compares the results of simulations with $\Q=1.5,7.5,15$ at fixed $\chi=0.02$, and with no chiral flipping. One can see that the model with $\Q=7.5$ follows similar quenching dynamics as the fiducial simulation with $\Q=15$. The chirality imbalance rises to $\tmu\approx 4$, and then falls to $\tmu\approx 0.3$ for $\tt\gtrsim 4$. This is accompanied by a rise in $\lambda H$, which then grows linearly until the end of the driving phase at $\tt=\Q$ in each simulation. The growth of seed fields and saturation of the CPI requires time $\tt\sim 4$, and hence a minimum value of $\Q\sim 4$. The results confirm that our fiducial value of $\Q=15$ is sufficient to capture the quenching dynamics expected in the limit of $\Q\gg 1$ in the absence of chiral flipping, $\Gf=0$. Larger values of $\Q$ are needed to fully capture the dynamics with $\Gf\neq 0$, as shown in Section~\ref{sec:SimFlipping} using the simulation with $\Q=150$.

By contrast, the simulation with $\Q=1.5$ is similar to the behavior of models with $\Q\ll1$ examined in previous studies where the chirality imbalance reaches its maximum value $\mu_0$ (or, equivalently, $\tmu=\Q$) before the CPI can saturate. In this case, $\tmu$ rises to $\Q=1.5$ by $\tt=1.5$ and then stays approximately constant until the CPI eventually saturates, converting a large part of $\mu$ into $\lambda H$ at $\tt\gtrsim 9$.

\begin{figure}
    \centering
    \includegraphics[width=\linewidth]{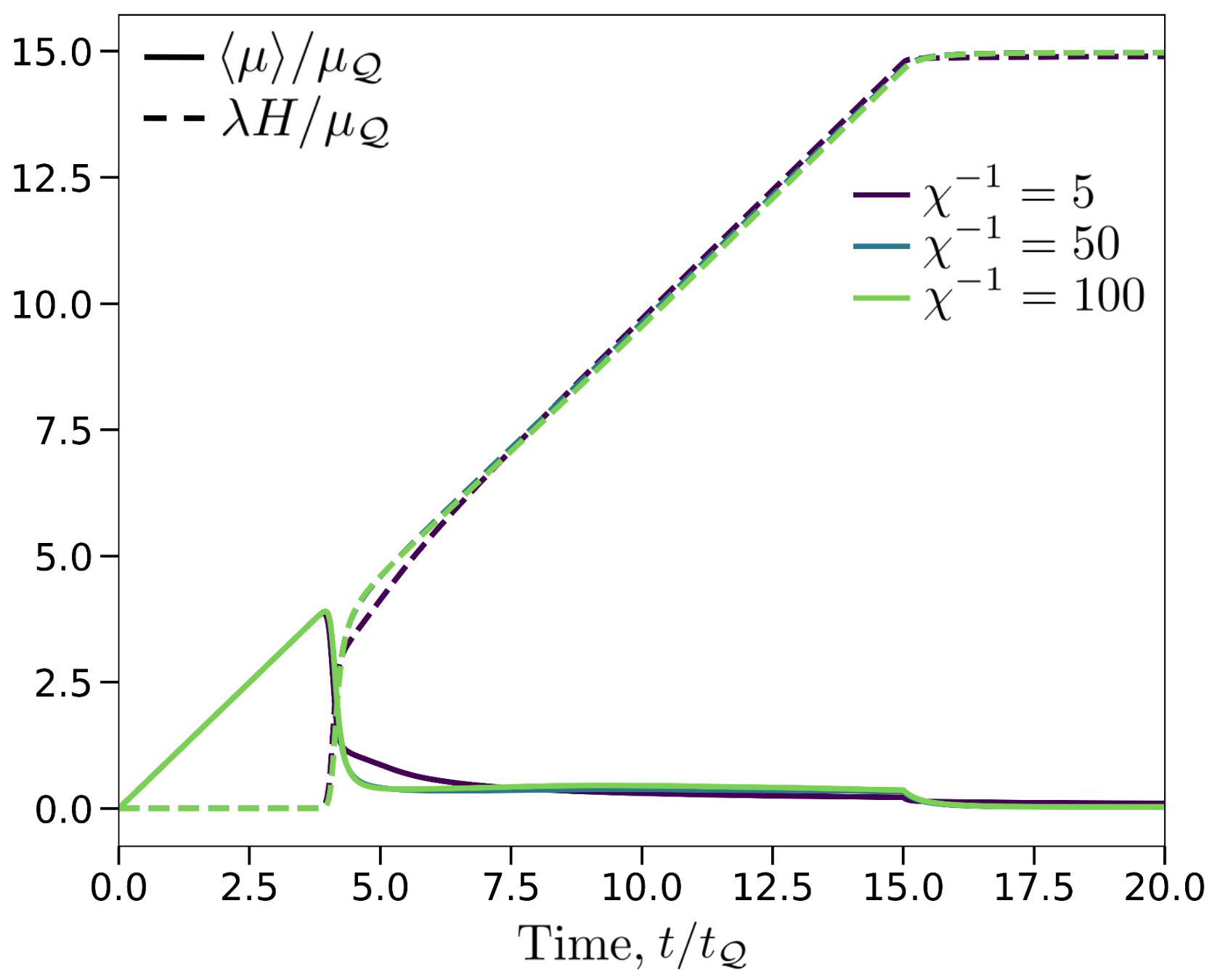}
    \caption{
    Comparison of the simulations with $\chi^{-1}=5,50$, and 100. All three simulations have $\Q=15$ and no chiral flipping, $\Gf=0$. Solid curves show the chirality imbalance $\langle \mu\rangle$ and dashed curves show the magnetic helicity $\lambda H$. The results are nearly identical for the large $\chi=50$ and 100 (the corresponding curves are indistinguishable in the figure).
    }
    \label{fig:Chi_scan}
\end{figure}

Fig.~\ref{fig:Chi_scan} compares models with $\chi^{-1}=5,50,100$ at fixed $\Q=15$. One can expect that the MHD turbulence driven by the CPI is near equipartition, $u\sim B/\sqrt{4\pi\rho}$, if $\chi^{-1}$ is large enough for the velocity and magnetic fields to couple efficiently. Simulation results are expected to be independent of $\chi$ if the Reynolds number $\Rm=\widetilde u_{\mathrm{rms}}\tL/4\chi$ is sufficiently large, $\Rm\gtrsim 10$. This expectation is confirmed in Fig.~\ref{fig:Chi_scan}: the results obtained at $\chi^{-1}=50$ and 100 are nearly identical. This verifies that our fiducial model with $\chi^{-1}=50$ is safely in the relevant astrophysical limit of $\Rm\gg1$.


\bibliography{apssamp}

\end{document}